\documentclass[11pt,a4paper]{article}
\usepackage{amssymb} \usepackage{amsmath} \usepackage{graphicx}
\usepackage{slashed}
\usepackage{epsfig,latexsym,color,xcolor}
\usepackage{ulem}
\usepackage{amstext}    
\usepackage{array} 
\usepackage{amsmath}
\begin{document}
\def\Giulia{\bf\color{red}}
\def\bef{\begin{figure}}
\def\eef{\end{figure}}
\newcommand{\ans}{ansatz }
\newcommand{\be}[1]{\begin{equation}\label{#1}}
\newcommand{\beq}{\begin{equation}}
\newcommand{\ee}{\end{equation}}
\newcommand{\beqn}[1]{\begin{eqnarray}\label{#1}}
\newcommand{\eeqn}{\end{eqnarray}}
\newcommand{\bd}{\begin{displaymath}}
\newcommand{\ed}{\end{displaymath}}
\newcommand{\mat}[4]{\left(\begin{array}{cc}{#1}&{#2}\\{#3}&{#4}
\end{array}\right)}
\newcommand{\matr}[9]{\left(\begin{array}{ccc}{#1}&{#2}&{#3}\\
{#4}&{#5}&{#6}\\{#7}&{#8}&{#9}\end{array}\right)}
\newcommand{\matrr}[6]{\left(\begin{array}{cc}{#1}&{#2}\\
{#3}&{#4}\\{#5}&{#6}\end{array}\right)}
\newcommand{\cvb}[3]{#1^{#2}_{#3}}
\def\lsim{\raise0.3ex\hbox{$\;<$\kern-0.75em\raise-1.1ex
e\hbox{$\sim\;$}}}
\def\gsim{\raise0.3ex\hbox{$\;>$\kern-0.75em\raise-1.1ex
\hbox{$\sim\;$}}}
\def\abs#1{\left| #1\right|}
\def\simlt{\mathrel{\lower2.5pt\vbox{\lineskip=0pt\baselineskip=0pt
           \hbox{$<$}\hbox{$\sim$}}}}
\def\simgt{\mathrel{\lower2.5pt\vbox{\lineskip=0pt\baselineskip=0pt
           \hbox{$>$}\hbox{$\sim$}}}}
\def\unity{{\hbox{1\kern-.8mm l}}}
\newcommand{\eps}{\varepsilon}
\def\ep{\epsilon}
\def\ga{\gamma}
\def\Ga{\Gamma}
\def\om{\omega}
\def\omp{{\omega^\prime}}
\def\Om{\Omega}
\def\la{\lambda}
\def\La{\Lambda}
\def\al{\alpha}
\def\beq{\begin{equation}}
\def\eeq{\end{equation}}
\newcommand{\ov}{\overline}
\renewcommand{\to}{\rightarrow}
\renewcommand{\vec}[1]{\mathbf{#1}}
\newcommand{\vect}[1]{\mbox{\boldmath$#1$}}
\def\tm{{\widetilde{m}}}
\def\mcirc{{\stackrel{o}{m}}}
\newcommand{\Dm}{\Delta m}
\newcommand{\dm}{\varepsilon}
\newcommand{\tanb}{\tan\beta}
\newcommand{\nbar}{\tilde{n}}
\newcommand\PM[1]{\begin{pmatrix}#1\end{pmatrix}}
\newcommand{\up}{\uparrow}
\newcommand{\down}{\downarrow}
\newcommand{\refs}[2]{eqs.~(\ref{#1})-(\ref{#2})}
\def\omE{\omega_{\rm Ter}}
\newcommand{\eqn}[1]{eq.~(\ref{#1})}
%

\newcommand{\DSUSY}{{SUSY \hspace{-9.4pt} \slash}\;}
\newcommand{\DCP}{{CP \hspace{-7.4pt} \slash}\;}
\newcommand{\mc}{\mathcal}
\newcommand{\gr}{\mathbf}
\renewcommand{\to}{\rightarrow}
\newcommand{\gtc}{\mathfrak}
\newcommand{\wh}{\widehat}
\newcommand{\br}{\langle}
\newcommand{\kt}{\rangle}


\def\lsim{\mathrel{\mathop  {\hbox{\lower0.5ex\hbox{$\sim$}
\kern-0.8em\lower-0.7ex\hbox{$<$}}}}}
\def\gsim{\mathrel{\mathop  {\hbox{\lower0.5ex\hbox{$\sim$}
\kern-0.8em\lower-0.7ex\hbox{$>$}}}}}

\def\nn{\\  \nonumber}
\def\de{\partial}
\def\brf{{\mathbf f}}
\def\bbf{\bar{\bf f}}
\def\bF{{\bf F}}
\def\bbF{\bar{\bf F}}
\def\bA{{\mathbf A}}
\def\bB{{\mathbf B}}
\def\bG{{\mathbf G}}
\def\bI{{\mathbf I}}
\def\bM{{\mathbf M}}
\def\bY{{\mathbf Y}}
\def\bX{{\mathbf X}}
\def\bS{{\mathbf S}}
\def\bb{{\mathbf b}}
\def\bh{{\mathbf h}}
\def\bg{{\mathbf g}}
\def\bla{{\mathbf \la}}
\def\bmu{\mathbf m }
\def\by{{\mathbf y}}
\def\bmu{\mbox{\boldmath $\mu$} }
\def\bsig{\mbox{\boldmath $\sigma$} }
\def\bunity{{\mathbf 1}}
\def\cA{{\cal A}}
\def\cB{{\cal B}}
\def\cC{{\cal C}}
\def\cD{{\cal D}}
\def\cF{{\cal F}}
\def\cG{{\cal G}}
\def\cH{{\cal H}}
\def\cI{{\cal I}}
\def\cL{{\cal L}}
\def\cN{{\cal N}}
\def\cM{{\cal M}}
\def\cO{{\cal O}}
\def\cR{{\cal R}}
\def\cS{{\cal S}}
\def\cT{{\cal T}}
\def\eV{{\rm eV}}

\numberwithin{equation}{section}

\vspace{6mm}

\large
 \begin{center}
 {\Large \bf  
Tests of Pauli Exclusion Principle Violations from Non-commutative quantum gravity}

 \end{center}

 \vspace{0.1cm}



\begin{center}
{\large Andrea Addazi}\footnote{E-mail: \, andrea.addazi@qq.com}
\\
{\it Department of Physics \& Center for Field Theory and Particle Physics, Fudan University, 200433 Shanghai, China}

\end{center}

\begin{center}

{\large Rita Bernabei}\footnote{E-mail: rita.bernabei@roma2.infn.it}
\\
{\it INFN sezione Roma ``Tor Vergata", I-00133 Rome, Italy, EU; 

Dipartimento di Fisica, Universit\`a di Roma ``Tor Vergata", I-00133 Rome, Italy, EU.}

\end{center}

\vspace{1cm}
\begin{abstract}
\large
\noindent 
\noindent 
We review the main recent progresses in non-commutative space-time phenomenology in underground experiments. 
A popular model of non-commutative space-time is $\theta$-Poincar\'e model, based on the 
Groenewold-Moyal plane algebra. This model predicts a violation of the Spin-statistic theorem,
in turn implying an energy and angular dependent violation of the Pauli Exclusion principle. 
Pauli Exclusion Principle Violating transitions in nuclear and atomic systems 
can be tested with very high accuracy in underground laboratory experiments 
such as DAMA/LIBRA and VIP(2). 
In this paper we derive that the $\theta$-Poincar\'e model can be already 
ruled-out until the Planck scale, from nuclear transitions tests by DAMA/LIBRA experiment. 

\end{abstract}

\baselineskip = 20pt

\section{Introduction}


It is a quite common believing that no any bounds to quantum gravity effects may be provided from next future experiments. 
The energy-scales probed by current and future collider experiments are far below the Planck scale. 
It is worth to remind that the Large Hadron Collider (LHC) tests energy scales of about $1-10\, {\rm TeV}$ or so, i.e. 
 15th $-$ 16th order of magnitude down to the Planck energy scale. 
 Certainly, this may inspire a certain pessimism to any serious attempts of quantum gravity phenomenology.

However, new recent progresses opened the way to a new exciting possibility building a bridge 
from experiments to quantum gravity physics. 
In Ref.~\cite{Addazi:2017bbg,Addazi:2018jmt}, we propose to search 
for  exotic transitions in nuclei or atoms, induced by non-commutative space-time, which violates the Pauli Exclusion Principle. 
Certainly, a possible detection of a PEP violating transition has 
the wonderful potentiality to change our conceptions of space and time. 
The Pauli Principle is a direct consequence of the Spin Statistic theorem (SST), in the Standard Model of particle physics.
 In turn, SST is valid under assumptions of: Minkowski's space-time, causality, locality and the Poincar\'e symmetry group. 
 The detection of Pauli Exclusion Principle Violations (PEPV) may be lead to indirect quantum gravity smoking guns in underground experiments of rare processes physics. 
 A non-commutative quantum gravity model related to  PEPV transitions is the $\theta$-Poincar\'e. 
The $\theta$-Poincar\'e is based on a deformation of the Poincar\'e symmetry.
This model entails a dual reformulation, in terms of non-commutative space-time coordinates. 

The $\theta$-Poincar\'e co-algebra can be obtained from the Poincar\'e algebra thanks to a mathematical map, known as the Groenewold-Moyal (GM) map
\cite{Majid:1996kd,Oeckl:2000eg,Chaichian:2004za,Aschieri:2005yw}. 
The same GM product map is applied to every quantum field theory operators such as 
creation/annihilation particle operators and every fields (electro-weak, chromo-strong and Higgs fields). In other words, a deformed version of the Standard Model of particle physics may be obtained as a Groenewold-Moyal Standard Model (GSSM). In GSSM, one can easily obtain all GM Feynman diagrams from the standard ones. However, most of the amplitudes are corrected by harmonic functions, which are dependent on the particles four-momenta.


In $\theta$-Poincar\'e model, the Poincar\'e algebra is deformed by the GM map as follows.
The space-time translations $x^{\mu}\rightarrow x^{\mu}+a^{\mu}$ 
are undeformed by GM map: 
$${\rm translation} \rightarrow {\rm translation}\, . $$

The action on the Lorentz group --- namely $\mathfrak{so}(3,1)$ --- generators is less trivial: 
$$\,\mathfrak{so}(3,1)\rightarrow {\rm ``deformed''}\,\,\, \mathfrak{so}(3,1)\,. $$

 As aforementioned, not only space-time generators, but also quantum operators related to particle fields are deformed as follows: 
$$\, {\rm (creation/annihilation\, ops.)} \rightarrow {\rm (GM-phase)(creation/annihilation\, ops.)\rm}\,,$$
$$\,{\rm (fields)} \rightarrow {\rm (GM-phase)(fields)}\,,$$

The GM provides a non-ambiguous map among the standard second quantization in the Standard Model
and the quantization in $\theta$-Poincar\'e.  \\

The most interesting aspect of $\theta$-Poincar\'e is that it is an example of quantum gravity model 
which, surprisingly, not only can be tested, but, even more surprisingly, is already ruled out in its {\it democratic} implementation by several underground experiments' data \cite{Addazi:2017bbg,Addazi:2018jmt}. The $\theta$-Poincar\'e models can be distinguished in two classes: the {\it democratic} and the {\it despotic} cases. 
The {\it democratic} $\theta$-Poincar\'e models assume that all the Standard Model fields 
interact with the non-commutative space-time background with the same gravitational coupling,
while the {\it despotic} case relaxes such a hypothesis.  
The main point is that $\theta$-Poincar\'e can induce very tiny but testable Pauli forbidden transitions \cite{Addazi:2017bbg}. Contrary to effective PEP violating 
models proposed in Refs.~\cite{Messiah:1900zz,Greenberg:1963kk,Gentile,Green:1952kp,Ignatiev:1987zd,Gavrin:1988nu,PEP1,PEP2,PEP3}, such transitions are: i) energy dependent from the particular PEPV process considered; ii) suppressed with the non-commutative energy scale; iii) highly motivated by quantum gravity.

In this review, we will show estimations of PEPV atomic/nuclear level transitions
induced by $\theta$-Poincar\'e. We will show how that underground experiments can rule out  $\theta$-Poincar\'e models up to non-commutative length scales, beyond the Planck scale. Rare processes, in nuclear and atomic physics, can provide indirect probes of the same structure of space and time. 
We will show how BOREXINO, KAMIOKANDE and DAMA still exclude $\theta$-Poincar\'e coupled to hadrons, 
beyond the Planck scale, as a phenomenological tombstone 
for the democratic scenario. 


\section{Atomic and Nuclear transitions}

Let us consider the one-particle state  
\be{alpha}
|\alpha\rangle =\langle a^{\dagger},\alpha |0\rangle=\langle c^{\dagger},\alpha |0\rangle=\int \frac{d^{d}p}{2p_{0}}\alpha(p)c^{\dagger}(p)\, ,
\ee
opportunely normalized as 
\be{alpha}
\langle\alpha|\alpha\rangle=1,\,\,\,\qquad  \int \frac{d^{d}p}{2p_{0}}|\alpha(p)|^{2}=1\, . 
\ee
From the definition in Eq.~(\ref{alpha}), we may construct a two-identical-particles state that, in $\theta$-Poincar\'e, reads  
\be{two}
|\alpha,\alpha\rangle = \langle a^{\dagger},\alpha\rangle  \langle a^{\dagger},\alpha\rangle  |0\rangle\, =
\ee
$$=\int \frac{d^{d}p_{1}}{2p_{10}}\frac{d^{d}p_{2}}{2p_{10}}e^{-\frac{\imath}{2}p_{1\mu}\theta^{\mu\nu} p_{2\nu}}\alpha(p_{1}) \alpha(p_{2})c^{\dagger}(p_{1})c^{\dagger}(p_{2})|0\rangle\,,$$
where the $e^{-\frac{\imath}{2}p_{1\mu}\theta^{\mu\nu} p_{2\nu}}$ provides the GM deformation to the Standard Model case (for $\theta\rightarrow 0$ we reobtain the Standard two particle state). 
The two particle state must be normalized with the following norm: 
\be{alphaapha}
N=\langle\alpha,\alpha|\alpha,\alpha \rangle=\int \frac{d^{d}p_{1}}{2p_{10}}\frac{d^{d}p_{2}}{2p_{20}}(\bar{\alpha}(p_{1})\alpha(p_{1}))(\bar{\alpha}(p_{2})\alpha(p_{2}))(1-e^{-\imath p_{1\mu}\theta^{\mu\nu} p_{2\nu}})
\ee
$$
=\int \frac{d^{d}p_{1}}{2p_{10}}\frac{d^{d}p_{2}}{2p_{20}}(\bar{\alpha}(p_{1})\alpha(p_{1}))(\bar{\alpha}(p_{2})\alpha(p_{2}))(1-\cos( p_{1\mu}\theta^{\mu\nu} p_{2\nu}))\,.$$
We can redefine the two-particles state as follows: 
\be{norm}
|\alpha,\alpha\rangle \rightarrow  \frac{1}{N(\alpha,\alpha)}|\alpha,\alpha\rangle,\,\,\qquad \, \langle \alpha|\alpha\rangle=1\, .
\ee

Now let us come to the crucial point: the transition amplitude for the overlap probability that a two-different-particles state evolves into a two-identical-particles state. In the case of fermions, the amplitude is as follows: 
\be{alphalpa}
\langle \beta,\gamma|\alpha, \alpha\rangle=\frac{1}{N}\int \frac{d^{d}p_{1}}{p_{10}}\frac{d^{d}p_{2}}{p_{20}}(\bar{\beta}(p_{1})\alpha(p_{1}))(\bar{\gamma}(p_{2})\alpha(p_{2}))\Big[1-e^{-\imath p_{1\mu}\theta^{\mu\nu} p_{2\nu}} \Big]
\ee
$$= \frac{1}{N}\int \frac{d^{d}p_{1}}{p_{10}}\frac{d^{d}p_{2}}{p_{20}}(\bar{\beta}(p_{1})\alpha(p_{1}))(\bar{\gamma}(p_{2})\alpha(p_{2}))\Big[1-\cos\Big(p_{1\mu}\theta^{\mu\nu} p_{2\nu}\Big) \Big]\,.$$

It is trivial to check that, for $\theta\rightarrow 0$, the overlap amplitude vanishes out. But, for $\theta \neq 0$, the Pauli principle is violated, if the states are composed of fermions.
A two-fermions state has a non zero probability to transit into a state in which fermions are identical.

 Let us consider indeed the GM effective Hamiltonian density, which is expressed by  
\be{HHH}
H_{GM,ij}=\langle \Psi_{i}^{\theta}|V_{\theta}|\Psi_{j}^{\theta}\rangle=\langle \Psi_{i}^{0}|\mathcal{H}_{E}|\Psi_{j}^{0}\rangle = V_{0}\Big\{\cos (\phi_{PEPV})-\cos\Big(\phi_{PEPV}+p_{1\mu}\theta^{\mu\nu} p_{2\nu}\Big)\Big\}\,,
\ee
and for a central interaction potential, 
\be{central}
2\phi_{PEPV}=p_{1}\wedge p_{2}-p_{1}'\wedge p_{2}'-p_{1}'\wedge p_{1}+p_{2}'\wedge p_{2}\,,
\ee
where $p\wedge q=p^{\mu}\theta_{\mu\nu}q^{\nu}$. 
The PEPV phases are provided both by Eq.~\eqref{alphalpa} and the central interaction potential,
where $p_{1,2}$ are the initial momenta of interacting particles while $p_{1,2}'$ are the out ones.

Now, let us consider the problem of atomic level transitions.
In this case, we can consider a non-relativistic quantum mechanics approach, based on 
perturbation theory. The effective Hamiltonian is the 0th order standard one 
plus a PEPV perturbation term:  
$$H=H_{0}+V_{I,0}+V_{I,0}\phi_{PEPV}^{2}\, . $$
The 1st order perturbation coefficient is
\be{ckkk}
\dot{c}_{b}^{(1)}(t)=(\imath \hbar)^{-1}H_{ba}'(t)e^{\imath \omega_{ba}t}\,.
\ee
If the perturbation is time independent, we obtain 
\be{cbb}
c_{b}^{(1)}(t)=-\frac{H'_{ba}}{\hbar \omega_{ba}}(e^{\imath \omega_{ba}t}-1). 
\ee
The transition probability is, then, found to be 
\be{PPP}
P_{ba}(t)=|c_{b}^{(1)}(t)|^{2}=\frac{2}{\hbar}|H'_{ab}|^{2}F(t,\omega_{ba})=\frac{2}{\hbar}V_{0}^{2}\phi^{2}F(t,\omega_{ba})\,,
\ee
with $F=(1-\cos \omega t)/\omega^{2}$. In the long-time adiabatic approximation, Eq.(\ref{PPP}), by means of $F\rightarrow \pi t \delta(\omega)$, leads to  
\be{WWW}
W=\frac{2\pi}{\hbar}|H'_{ba}|^{2}=\frac{2}{\pi \hbar}V_{0}^{2}\phi^{2}_{PEPV}=W_{0}\phi^{2}_{PEPV}\,.
\ee
 However, such a quartic power suppression in $\theta$ powers may be an artifact, of the number of fields involved in the initial state. In presence of three particles, Eq.~(\ref{alphalpa}) 
 leads to a linear order correction in the phase $\phi_{PEPV}$. In this latter case we obtain
\be{WWW}
W\simeq W_{0}\phi_{PEPV}\, ,
\ee
where 
$\phi_{PEPV}$ coincides with the $\delta^{2}$ parameter, parametrizing the PEP deviation from creation/annihilation commutators,  constrained by experimental measurements.

Now we can distinguish two cases, corresponding to different choices of the $\theta$-components. In the first case, the time-space (electric) components is set to
zero:
\be{thetazi}
\theta_{0i}=0\rightarrow \phi_{PEPV}=\frac{1}{2}\Big(p_{1}^{i}\theta_{ij} p_{2}^{j}-p_{1}'^{i}\theta_{ij} p_{2}'^{j}-p_{1}'^{i}\theta_{ij} p_{1}^{j}+p_{2}'^{i}\theta_{ij} p_{2}^{j}\Big)\,.
\ee
Let us consider particle $1$ as an electron and the particle $2$ as a nucleus. Then all terms involving $p_{1}$ and $p_{1}'$ are subleading to $p_{2},p_{2}'$, while $|p_{2}|$ and 
$|p_{2}'|$ are of the order of the energy levels in the atom. 
Therefore, $p_{1}\wedge p_{1}'$ and $p_{2}\wedge p_{2}'$ are subdominant (the first for magnitude subdominance $|p_{1}|,|p_{1}'|<<|p_{2}|,|p_{2}'|$,
the second for $p_{2}\simeq p_{2}'$). Therefore, the relevant terms are $p_{1}^{i}\theta_{ij} p_{2}^{j}-p_{1}'^{i}\theta_{ij} p_{2}'^{j}$,
which is $|p_{1}|\hat{{\bf p}}_{1}\cdot \theta \cdot |p_{2}|\hat{{\bf p}_{2}}-|p_{1}'|\hat{{\bf p}}_{1}'\cdot \theta \cdot |p_{2}'|\hat{{\bf p}_{2}'}$. 
Introducing the cutoff UV dimension energy $\Lambda$ hidden in the dimensionful $\theta_{\mu\nu}$ 
and redefining $\theta$ as a antisymmetric dimensionless tensor; the GM-Standard Model predicts the result as follows
\be{EEEE}
\phi_{PEPV}\simeq \frac{1}{2} C \frac{\bar{E}_{1}}{\Lambda}\frac{\bar{E}_{1}'}{\Lambda}\,,
\ee
where $\bar{E}_{1},\bar{E}_{1}'$ are the energy levels occupied by the initial and the final electrons, while $C=\hat{{\bf p}}_{1}\cdot \theta \cdot \hat{{\bf p}}_{2}$.
The PEPV phase as an angular function is displayed in Fig. 1.

The second case has an extra phase with respect to the first one as follows: 
\be{thetazi}
\theta_{0i}\neq 0\rightarrow \Delta \phi_{PEPV}=\frac{1}{2}\Big(p_{1}^{0}\theta_{0j} p_{2}^{j}-p_{1}'^{0}\theta_{0j} p_{2}'^{j}-p_{1}'^{0}\theta_{0j} p_{1}^{j}+p_{2}'^{0}\theta_{0j} p_{2}^{j}\Big)+(0\leftrightarrow j)\,,
\ee
with
\be{PEPVphi}
\phi_{PEPV}\simeq  \frac{D}{2}\frac{E_{N}}{\Lambda}\frac{\Delta E}{\Lambda}   \, , 
\ee
where $E_{N}\simeq m_{N}\simeq A m_{p}$ is the nuclear energy, and $\Delta E=E_{1}'-E_{1}$ is the transition energy of the electron.


\begin{figure}[t]
\centerline{ \includegraphics [height=8cm,width=0.5\columnwidth]{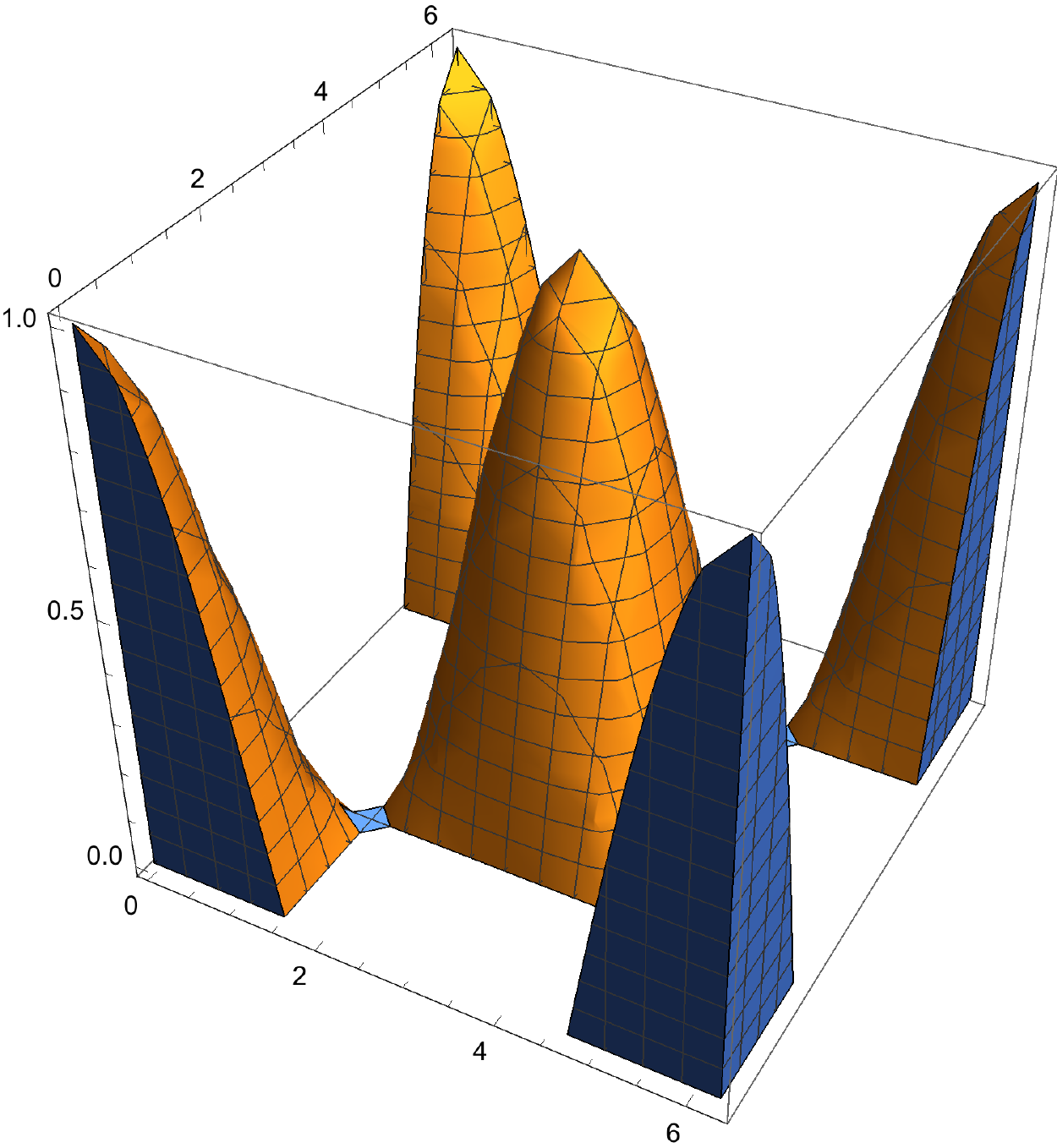}}
\vspace*{-1ex}
\caption{ Three-dimensional plot of the PEPV phase as a function of the angles (in x,y axes) between momenta and ${\bf \theta}$-matrix -- for fixed characteristic scales normalized to $1$. Here the plot for the
magnetic-like case is shown.  }
\label{plot}   
\end{figure}

\noindent 

\section{Pauli violating transitions in underground experiments: atomic and nuclear processes } \label{mico}

\noindent

Let us discuss in the following  the phenomenological implications of PEPV in several underground experiments. 

\begin{figure}[t]
\centerline{ \includegraphics [height=13cm,width=1.3\columnwidth]{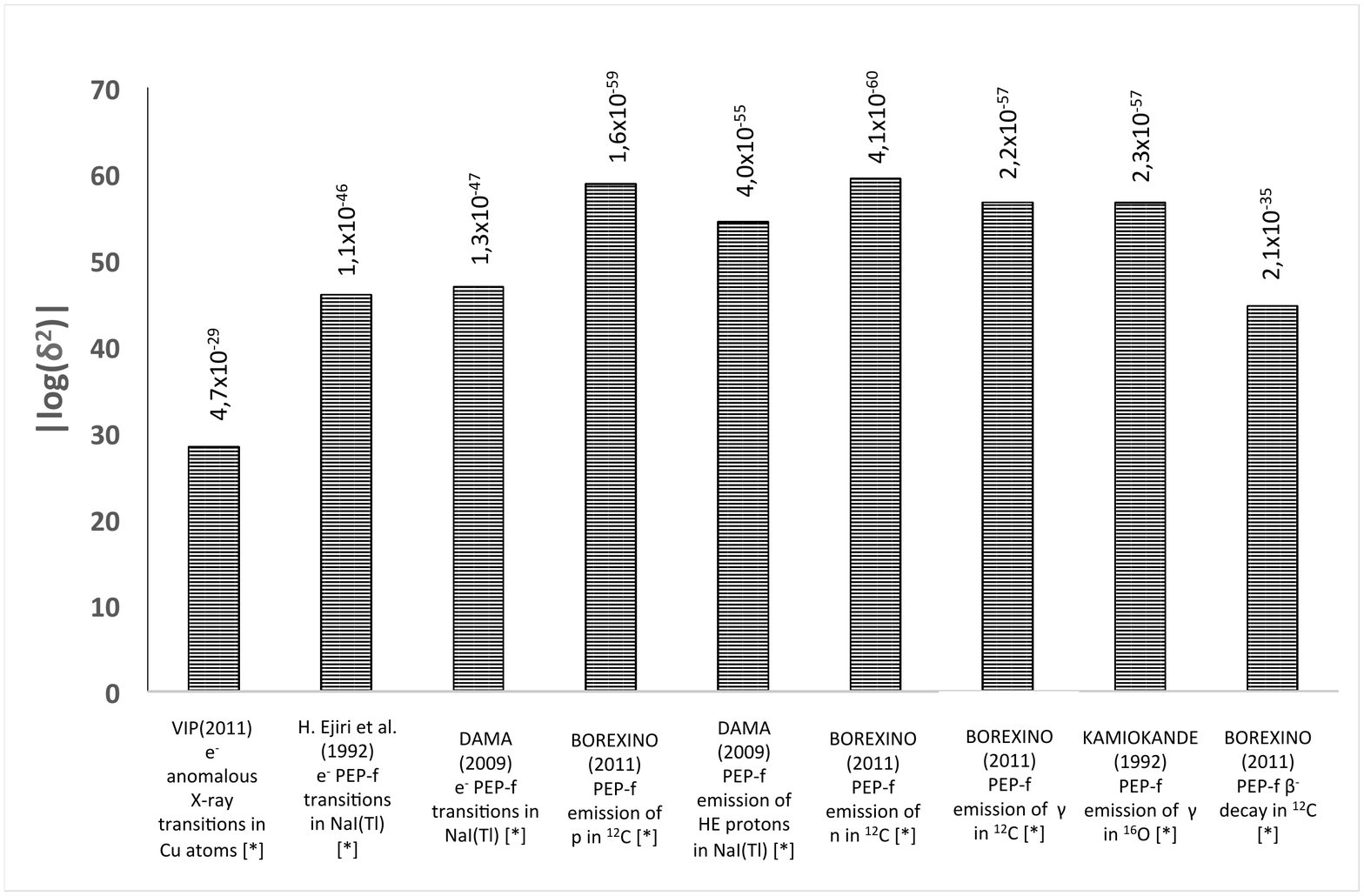}}
\vspace*{-1ex}
\caption{Limits at 90\% C.L. on various PEP violation channels in logarithmic scale, displaying the $-\log \delta^2$ as measured by various experimental collaborations: VIP(2011) \cite{Pichler:2016xqc}; ELEGANTS V (1992) \cite{Ejiri:1992}; DAMA(2009)A \cite{Bernabei:2009zzb}; MALBEK(2016) \cite{Abgrall:2016wtk};
Borexino(2011)A \cite{Bellini:2010}; DAMA(2009)B \cite{Bernabei:2009zzb}; Borexino(2011)B \cite{Bellini:2010}; Borexino(2011)C \cite{Bellini:2010}; Kamiokande(1992) \cite{Suzuki:1993zp}; Borexino(2011)D \cite{Bellini:2010} (Figure taken from Ref. 1. }
\label{plot}   
\end{figure}

\noindent 
 In Fig.2, we show limits on the PEPV strength ($\phi_{PEPV}=\delta^2$) parameter determined by the searches for new exotic transition [1]. 
In the following, we will recall various PEP experimental results provided by different experimental techniques. 

The VIP experiment searches for PEP violating atomic transitions in copper atoms \cite{Pichler:2016xqc}. 
The experimental technique is based on the injection of ``fresh" electrons into a copper material,
from circulating current. Possible PEP forbidden transitions are searched
 for by injecting \" fresh"
electrons into a copper strip and  searching for the X-rays following such forbidden radiative transitions occuring when one of these electrons is captured by a copper atom
and cascades down to the already-filled 1S state.
The energy gap corresponds to $\Delta E_{2P\rightarrow 1S}=7.729\, {\rm keV}$;
it should be compared with the 
ordinary $K_\alpha$ transition energy ($8.040\, {\rm keV}$).

Tests of
PEP forbidden electromagnetic atomic transitions, 
in Iodine atoms deploying NaI(Tl) detectors, have been performed by ELEGANTS V \cite{Ejiri:1992}  and DAMA/LIBRA  \cite{Bernabei:2009zzb} experiments. PEPV electromagnetic transitions
 in Germanium atoms in PPC HPGe detectors were searched for by the MALBEK experiment \cite{Abgrall:2016wtk}. 
 These experiments exploited a different strategy than VIP: 
PEPV transitions emit X-rays and Auger electrons, directly by the transition itself and by the following arrangements of the atomic shell. 
Very high detection efficiency, almost $100\%$, is achieved 
in the DAMA/LIBRA detectors; in particular, the whole  ionization energy for the considered shell is detected,  shifted by a certain
 $\Delta E$ related to the other electrons filling the shells.
The atomic K-shell provides the largest available energy emissions of X-rays  or Auger-electrons; 
however, severe limits, from DAMA/NaI, can be achieved also for L-shell transitions ($4\div 5$ keV radiation emission) in Iodine atoms \cite{bernabei2}
thanks to the low energy thresholds of the DAMA/NaI detectors.

It is worth noting that 
the most stringent constraints on PEPV, in atomic transitions, are provided by the DAMA/LIBRA experiment, 
searching for PEPV K-shell transitions in Iodine.
DAMA/LIBRA  consists of an about 250 kg array of highly radiopure NaI(Tl) detectors, hosted in the Gran Sasso National Laboratory (LNGS). 
The data set corresponds to 0.53 ton$\times$yr, implying  
a  limit on the PEPV transition characteristic time of $4.7 \times 10^{30}$ s.
This limit corresponds to 
  $\phi_{PEPV}=\delta^2 < 1.28 \times 10^{-47}$ at 90\% C.L. \cite{Bernabei:2009zzb}. This entails very strong constraints on the non-commutativity scale.
In the magnetic-like $\theta$-Poincar\'e scenario, $\Lambda < 10^{18}\, {\rm GeV}$ is excluded.
In the electric-like phase, the limit is less stringent than the magnetic-like case, but still arriving to very high energy scale: $\Lambda >5\times 10^{16}\, {\rm GeV}$. 

On the other hand, the most stringent bounds arrived from nuclear transitions, where the statistics can be even higher then atomic ones.

DAMA/LIBRA collaboration also sets severe limits on PEPV nuclear transitions \cite{Bernabei:2009zzb}. 
PEPV processes
 in nuclear shells of $^{23}$Na and $^{127}$I are investigated, emitting protons with an energy of E$_p \ge$ 10 MeV: 
the emission rate of protons with energy E$_p \ge$ 10 MeV from PEPV transitions, in $^{23}$Na and $^{127}$I, was constrained up to $\gsim 1.63 \times 10^{33}$ $s$ (90\% C.L.) \cite{Bernabei:2009zzb}, which 
corresponds to $\phi_{PEPV}=\delta^2 \lsim 4 \times 10^{-55}$ (90\% C.L.). 
Such a strong bound rules out 
 both the electric and the magnetic like $\theta$-Poincar\'e models with a non-commutative scale at the Planck scale energy.

\section{Conclusions and remarks} \label{co}


In this paper, aspects of Pauli Exclusion Violating processes induced by non-commutative $\theta$-Poincar\'e quantum gravity 
in underground experiments, are reviewed. 
In the following the main conclusions are summarized: 

\begin{itemize}
\item Pauli Violating transitions are sharp predictions of
 non-commutative $\theta$-Poincar\'e, i.e of the Groenewold-Moyal Standard Model (GMSM).

\item Predicted PEPV transitions are energy and angular dependent! In particular, they depend on the momenta
of the particles involved in the process. 

\item The PEPV Democratic scenario is already ruled out by DAMA/LIBRA experiment. The PEPV Despotic scenario, 
where non-commutativity is particle species dependent will be tested in atomic channels by VIP(2) experiments. 

\item Detectors with anisotropic response may also test the angular dependence of PEPV transitions. 
For example, a tempting possibility that we would like to suggest is that a good candidate for such 
searches is provided by \cite{Bernabei:2017yhw}, which was suggested for measuring the dark matter directionality 
(highly motivated by dark matter candidates as Mirror matter \cite{Addazi:2015cua,Cerulli:2017jzz}).

\end{itemize}

\vspace{0.5cm} 

{\large \bf Acknowledgments} 

\vspace{0.5cm}

We thank Pierluigi Belli, Catalina Oana Curceanu and Antonino Marciano
 for useful discussions and remarks on this subject. 
 AA would like to thank Pisin Chen and National Taiwan University (NTU) for hospitality during the preparation of this paper. 
AA acknowledges support by the NSFC, through the grant No. 11875113, the Shanghai Municipality, 
through the grant No. KBH1512299, and by Fudan University, through the grant No. JJH1512105.



\end{document}